\documentclass{eas}
\usepackage{graphicx}
\def\sbr{${\rm mag\,\,arcsec^{-2 }}$\ }

\begin{document}

\title{Faint dwarf galaxies in the Next Generation Virgo cluster Survey} 
\runningtitle{Dwarfs in the NGVS}

\author{Pierre-Alain Duc}\address{Laboratoire AIM Paris Saclay, CNRS/INSU, CEA/Irfu, Universit\'e Paris Diderot, France}
\author{Laura Ferrarese}\address{Herzberg Institute of Astrophysics, National Research Council of Canada, Canada}
\author{Jean-Charles Cuillandre}\address{Canada-France-Hawaii Telescope Corporation}
\author{Stephen Gwyn}\sameaddress{2}
\author{Lauren A. MacArthur}\sameaddress{2}
\author{Etienne Ferriere}\sameaddress{1}
\author{Patrick C\^ot\'e}\sameaddress{2}
\author{Patrick Durrell}\address{Department of Physics \& Astronomy, Youngstown State University, United--States}
\begin{abstract}
The Next Generation Virgo Cluster Survey (NGVS) is a CFHT Large Program that is using the wide field of view capabilities of the MegaCam camera to map the entire Virgo Cluster from its core to virial radius. The observing strategy has been optimized to detect very low surface brightness structures in the cluster, including intracluster stellar streams and faint dwarf spheroidal galaxies. We present here the current status of this ongoing survey, with an emphasis on the detection and analysis of the very low-mass galaxies in the cluster that have been revealed by the NGVS. 
\end{abstract}
\maketitle
\section{A new generation imaging survey of the Virgo Cluster}
\label{intro}
The Virgo Cluster, the closest large concentration of galaxies in the nearby Universe, has long been one of the favorite playgrounds of astronomers. From the 32 cluster members identified by Smith (\cite{Smith36}),  to the 2096 objects in the Virgo Cluster Catalog (VCC)  (Binggeli \etal\ \cite{Binggeli85b}), Virgo has provided large samples of galaxies of all morphological  types and masses --- key data to study the structure and evolution of galaxies. 
The VCC, despite its age --- the survey is based on photographic plates obtained with the Las Campanas Dupont 2.5m telescope between 1979 and 1982 ---  remains the reference catalog for any study of the Virgo Cluster.  More recently, the Sloan Digital Sky Survey (SDSS) has at last allowed Virgo to enter the digital age and colorized the cluster, bringing multi-wavelength shallow CCD images in 5 optical bands.   Deeper surveys exist for some specific areas of Virgo:  two E-W and N-S strips were mapped  with the 2.5m Isaac Newton Telescope at a depth allowing the detection of galaxies with central surface brightness fainter than 26~\sbr\  in the B band  (Sabatini \etal\ \cite{Sabatini03}; Roberts \etal\ \cite{Roberts07}).  Mihos \etal\ (\cite{Mihos05}) obtained with the 0.6m Burrell Schmidt Telescope   the deepest images yet available towards the most massive galaxies in the cluster, detecting new intracluster stellar streams and fine-structures  (Janowiecki \etal\  \cite{Janowiecki10}).

The Next Generation Virgo Cluster Survey (NGVS)\footnote{https://www.astrosci.ca/NGVS/Home.html}  aims to reach the sensitivity of the Burrell Survey, with excellent image quality  over the full area of the VCC survey, and providing all the colors of the SDSS. When completed, it will be the reference database for any optical study on the Virgo Cluster.
The NGVS is a Large Program of the 3.6m Canada-France-Hawaii-Telescope. It takes advantage of its excellent site  (the summit of the Mauna Kea), the wide field of view and good spatial resolution of the MegaCam camera, and the large number of hours allocated to the project (771 hours distributed over 4 semesters between 2009 and 2012).  
More specifically, NGVS provides a contiguous coverage in 5 bands ({\it u,g,r,i,z}) of the Virgo cluster from its core to its virial radius. The 104 deg$^2$ survey area (about 8.6 Mpc$^2$ at the distance of Virgo) is tiled in 117 slightly overlapping MegaCam pointings. 


\begin{figure}
\resizebox{1\columnwidth}{!}{
 \includegraphics{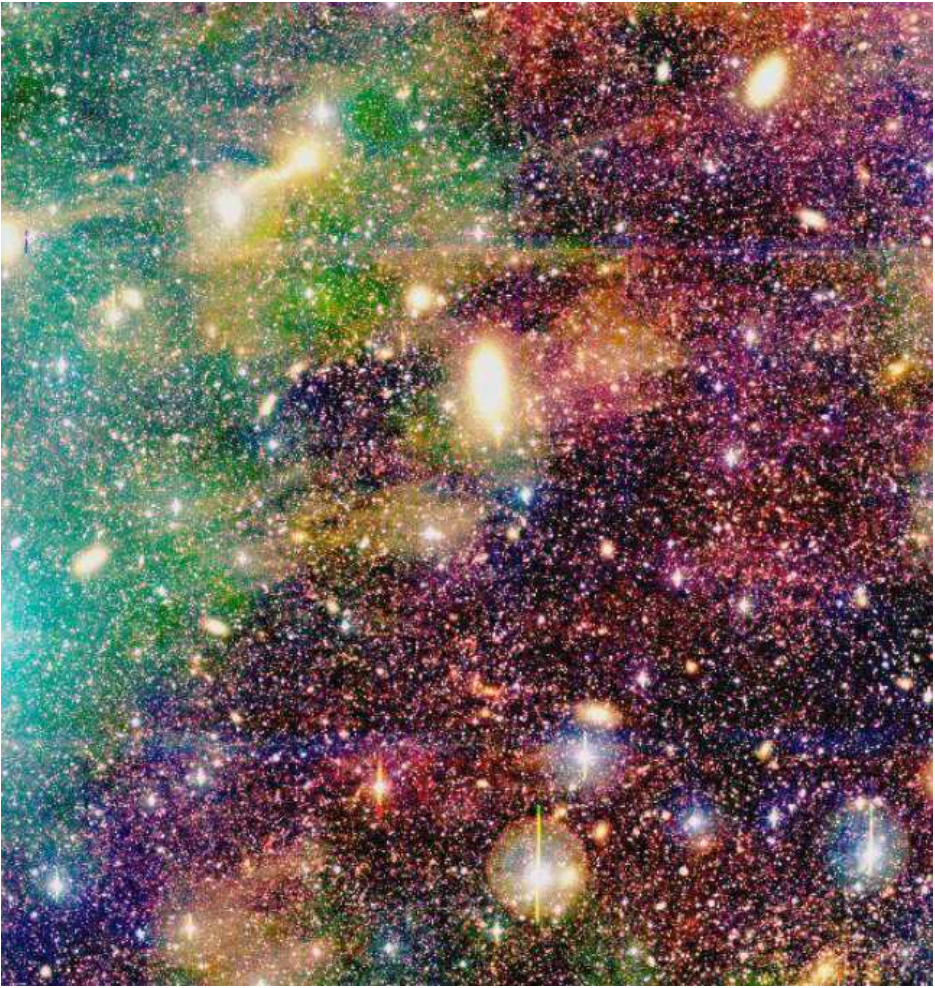} }
\caption{Composite image ({\it u}:blue, {\it g}:green,  {\it i}:red) for one of the NGVS fields, on a logarithm intensity scale. The low-surface brightness features are  emphasized, displaying the images   at their limiting surface brightness (29~\sbr in the {\it g}--band).}
\label{fig:crowded}       
\end{figure}

\section{Observational challenges}
\label{sec:obs}
Viewed from the NGVS, Virgo looks amazingly different than the vision of the cluster seen in some previous imaging studies. Fig.~\ref{fig:crowded} presents a composite image of one of the MegaCam fields. The cluster and its galaxies are hardly recognizable.  The  many features visible on this image are detailed below, ordered as a function of distance to the observer:

$\bullet$ instrumental signatures! MegaCam is a complex instrument hosting 36 individual CCD detectors  and numerous optical elements. In order to minimize their imprints on the final stacked images, specific observing strategies  and pipelines had to be implemented. A master sky is built and subtracted from all individual images before stacking them.  The pipeline {\it Elixir-LSB} developed at CFHT by J.-C. Cuillandre performs the data reduction, optimized for the detection of low-surface brightness structures. It allowed the reduction of  instrumental artifacts to a minimum: horizontal bands due to the lower sensitivity in the gaps between the CCDs are visible only when getting to  0.2\% of the sky level.

$\bullet$ a myriad of white dots, most of which are  faint  foreground stars.  Indeed, Virgo is located  behind the so-called Virgo over-density, a particularly crowded region of the Milky Way halo where several stellar streams cross (Law \etal\ \cite{Law05}). One of the aims of the NGVS is to study such streams which provide information about the formation of our own Galaxy. A fraction of the point-like sources are, in fact, intracluster globular clusters associated with Virgo galaxies.  The study of their spatial distribution --- see for instance the work by  Lee \etal\ (\cite{Lee10}) in the SDSS  ---  greatly facilitated by the high spatial resolution and depth of the NGVS, gives key information on the  dark matter profiles of the massive cluster galaxies as well as on the intracluster stellar streams.

$\bullet$  camemberts  crossed by spikes ... due to halos of bright foreground stars. The light around the most luminous stars  contaminates the low--surface brightness structures in the image and has to be carefully subtracted.

$\bullet$ diffuse green/yellow extended emission as well as parallel narrow patterns (especially visible to the East) due to foreground cirrus clouds in the Milky Way. Scattered emission from Galactic interstellar clouds and red emission associated with photoluminescence (Witt \etal\ \cite{Witt08}) can mimic stellar streams  (Cortese \etal\ \cite{Cortese10}). The detection of the intracluster light and remnants of past galactic  collisions -- one of the major goals of NGVS -- must therefore contend with the presence of cirrus. Even though they cannot be eliminated, they may be identified using ancillary data, in particular, from their far-infrared and ultraviolet emission. NGVS will benefit  from the availability of  parallel Herschel HeViCS (Davies \etal\ \cite{Davies10}) and GALEX GUViCS\footnote{http://galex.oamp.fr/guvics/} Virgo surveys.

$\bullet$ a few white elliptical stains... the galaxies themselves. Some of them are located behind Virgo such as the colliding system with prominent tidal features to the North-East of the image. Of course, this background science is a key component of the CFHT Large Program: providing a deep, contiguous field over an area of more than 100 square degrees, the NGVS allows searches for distant clusters of galaxies and extends the weak lensing analysis started with the CFHTLS,  one of the previous CFHT Large Programs dedicated to cosmology. 

One of the challenges facing the Next Generation Virgo Cluster Survey  is to identify ---  in the forest of  foreground stars, Galactic clouds and  background galaxies ---  objects that are actually located in the Virgo Cluster, in particular, the least massive ones that were discussed in depth during this conference. Our preliminary analysis, presented below,  indicates that this is indeed feasible and that it might even be possible to identify in Virgo counterparts of the (ultra)-faint dwarfs so far detected only in the Local Group.

\section{Dwarf galaxies in the NGVS}
             
There is a long history of studies of dwarf galaxies in Virgo, addressing the possible  evolutionary links between their various classes  -- dwarf  ellipticals, with or without nuclei, irregulars, blue compact, ultra-compact   (see for instance Lisker et al. in this volume) -- the effect of the environment  on their properties, or their scaling relation in connection with the more massive galaxies (Graham et al. in this volume; C\^ot\' e et al. 2010, submitted). 
  All these studies benefit from having a large sample of galaxies spanning the widest possible range in cluster-centric distance, such as that  provided by the NGVS. 
The other playground  of dwarf aficionados is the Local Group where recently  a new population of so-called ultra-faint dwarf galaxies has recently been discovered, studied and, in some cases, claimed that  they may (partly) solve the missing satellite problem in cosmology.   The NGVS allows for an extension of  the quest for the least massive galaxies to the distance of the Virgo Cluster. 

\begin{figure}
\resizebox{1\columnwidth}{!}{
  \includegraphics{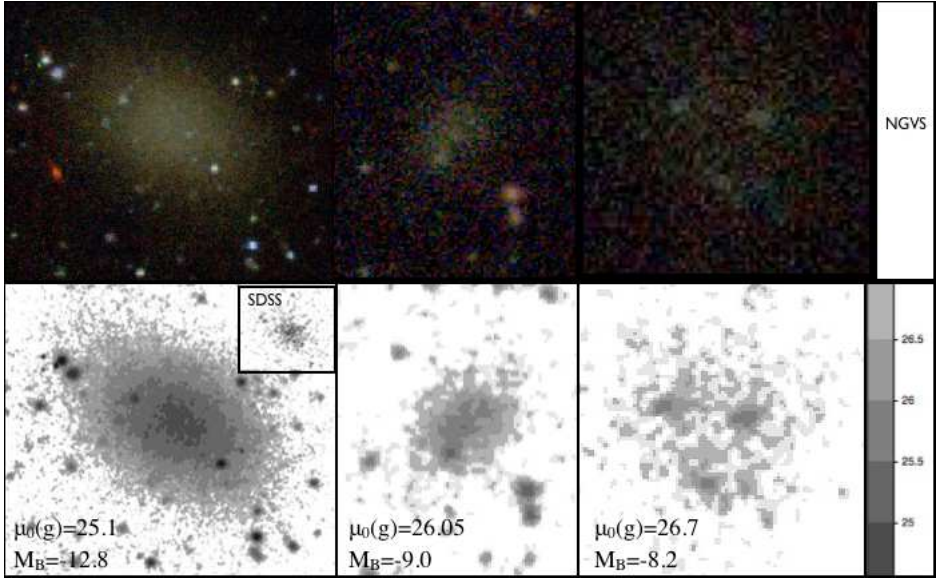} }
\caption{ Examples of three low-surface brightness dwarfs of various masses detected by the NGVS. {\it Top}: true color ({\it g,r,i}) images. {\it Bottom:} {\it g}--band surface brightness map, with the scale displayed to the right. The absolute blue magnitudes and central surface brightnesses in \sbr\ are indicated.   The most ``luminous" galaxy to the left, which belongs to the Virgo Cluster Catalog, is barely detected on the SDSS (see image in the inset). The two other galaxies are yet uncatalogued and undetected on the SDSS. The galaxy to the right, with an absolute blue magnitude of  -8.2, is close to the nominal dividing point between Local Group dSphs and ``ultra-faint" dwarf galaxies.}
\label{fig:ex}       
\end{figure}

\subsection{Detection}
The detection and photometry  of the NGVS  faint dwarfs  raises a number of technical issues and require software developments.

As a first step in these efforts, we have identified by eye Virgo LSB dwarf candidates in multi-band images of the central 4 square degrees of the cluster. This exercise was carried out by three independent groups and the results were cross-matched. 

In parallel, automatic pipelines are being developed. One code makes use of an optimized version of SExtractor run on ring-filtered images and the 2D profile fitting code Galfit (Peng \etal\ \cite{Peng02}). The second code benefits from the development of MARSIAA (MARkovian Software for Image Analysis in Astronomy) carried out at the Strasbourg University (Vollmer et al., 2010, submitted). This image segmentation software decomposes images into classes of pixels, each of these families being defined  by a specific  statistical behavior. One or several of these classes correspond to the LSB structure. MARSIAA processes simultaneously multi-band data images. An additional software, DetectLSB, digests the masks created by MARSIAA and identifies the LSB dwarfs  (see further details in the poster contribution of Ferriere et al. in this volume).
Each of these pipelines produces false-positive detections which should then be manually cleaned.


\subsection{Preliminary results}
The eye detection exercise performed in the central regions  led to the detection of $\approx$ 550 objects, the majority of which (about 360) were previously uncatalogued.  Based on their sizes, colors and surface brightnesses, they have a strong likelihood  of belonging to the cluster. Given their central surface brightness (up to 27 \sbr\ in the {\it g}-band), a spectroscopic confirmation  is far beyond the capabilities of current facilities.
Examples of true color images and surface brightness maps  for three Virgo LSB dwarfs are displayed in Fig.~\ref{fig:ex}. The absolute magnitude of the faintest one puts it close to the transition between dSphs and ultra-faint dwarfs in the Local Group.
The NGVS detection rate in the core --- about 140 objects per square degree --- is at least three times that achieved by  previous generations of deep surveys made towards specific  strips of the Virgo cluster (Roberts \etal\ \cite{Roberts07}).
The census will be extended to the full cluster when the automatic detection pipelines are fully functional.

\subsection*{Acknowledgments}
Many thanks to the SOC, LOC, and  to his chair, P. Prugniel, for a very enjoyable conference. These proceedings were written on behalf of the NGVS team.



%


\end{document}